\documentclass[aip,pop,amsfonts,amsmath,amssymb,superscriptaddress,reprint, floatfix, floats]{revtex4-1}

\usepackage{graphicx}
\usepackage{color}
\usepackage{dcolumn}

\newcommand{\etal}{{\it et\,al.}}
\newcommand{\unit}[1]{\,\mathrm{#1}}
\newcommand{\der}{\mathrm{d}}

\newcommand{\J}{\,\mathrm{J}}

\newcommand{\MeV}{\,\mathrm{MeV}}
\newcommand{\keV}{\,\mathrm{keV}}
\newcommand{\eV}{\,\mathrm{eV}}

\newcommand{\mm}{\,\mathrm{mm}}
\newcommand{\mum}{\,\mathrm{\mu m}}
\newcommand{\nm}{\,\mathrm{nm}}

\newcommand{\pcc}{\,\mathrm{cm^{-3}}}

\newcommand{\gcc}{\,\mathrm{g/cm^3}}

\newcommand{\ps}{\,\mathrm{ps}}

\newcommand{\fs}{\,\mathrm{fs}}

\newcommand{\Wcm}{\,\mathrm{W/cm^2}}

\newcommand{\kA}{\,\mathrm{kA}}

\newcommand{\kT}{\,\mathrm{kT}}
\newcommand{\T}{\,\mathrm{T}}

\newcommand{\degree}{^{\circ}}

\begin{document}

\title{Laser-driven strong magnetostatic fields with applications to charged beam transport and magnetized high energy-density physics} 

\author{J.J. Santos}
\email{joao.santos@u-bordeaux.fr}
\affiliation{Univ. Bordeaux, CNRS, CEA, CELIA (Centre Lasers Intenses et Applications), UMR 5107, F-33405 Talence, France}

\author{M. Bailly-Grandvaux}
\affiliation{Univ. Bordeaux, CNRS, CEA, CELIA (Centre Lasers Intenses et Applications), UMR 5107, F-33405 Talence, France}
\affiliation{Department of Mechanical and Aerospace Engineering, University of California at San Diego, La Jolla, CA 92093, USA}

\author{M. Ehret}
\affiliation{Univ. Bordeaux, CNRS, CEA, CELIA (Centre Lasers Intenses et Applications), UMR 5107, F-33405 Talence, France}
\affiliation{Institut f\"ur Kernphysik, Tech. Univ. Darmstadt, Germany}

\author{A.V. Arefiev}
\affiliation{Department of Mechanical and Aerospace Engineering, University of California at San Diego, La Jolla, CA 92093, USA}

\author{D. Batani}
\affiliation{Univ. Bordeaux, CNRS, CEA, CELIA (Centre Lasers Intenses et Applications), UMR 5107, F-33405 Talence, France}

\author{F.N. Beg}
\affiliation{Department of Mechanical and Aerospace Engineering, University of California at San Diego, La Jolla, CA 92093, USA}

\author{A. Calisti}
\affiliation{PIIM, Univ. Aix-Marseille - CNRS, France}

\author{S. Ferri}
\affiliation{PIIM, Univ. Aix-Marseille - CNRS, France}

\author{R. Florido}
\affiliation{iUNAT, Departamento de F\'isica, Univ. de Las Palmas de Gran Canaria, 35017 Las Palmas de Gran Canaria, Spain}

\author{P. Forestier-Colleoni}
\affiliation{Univ. Bordeaux, CNRS, CEA, CELIA (Centre Lasers Intenses et Applications), UMR 5107, F-33405 Talence, France}
\affiliation{Department of Mechanical and Aerospace Engineering, University of California at San Diego, La Jolla, CA 92093, USA}

\author{S. Fujioka}
\affiliation{Institute of Laser Engineering, Osaka University, 2-6 Yamada-oka, Suita, Osaka, 565-0871, Japan}

\author{M.A. Gigosos}
\affiliation{Departamento de F\'isica Te\'orica, At\'omica y \'Optica, Universidad de Valladolid, 44071 Valladolid, Spain}

\author{L. Giuffrida}
\affiliation{Univ. Bordeaux, CNRS, CEA, CELIA (Centre Lasers Intenses et Applications), UMR 5107, F-33405 Talence, France}
\affiliation{Institute of Physics AS\v{C}R, v.v.i. (FZU), ELI-Beamlines, Doln\'{i} B\v{r}e\v{z}ny, Czech Republic}

\author{L. Gremillet}
\affiliation{CEA, DAM, DIF, F-91297 Arpajon, France}

\author{J.J. Honrubia}
\affiliation{ETSI Aeron\'auticos, Universidad Polit\'ecnica de Madrid, Madrid, Spain}

\author{S. Kojima}
\affiliation{Institute of Laser Engineering, Osaka University, 2-6 Yamada-oka, Suita, Osaka, 565-0871, Japan}

\author{Ph. Korneev}
\affiliation{National Research Nuclear University MEPhI, 115409, Moscow, Russian Federation}

\author{K.F.F. Law}
\affiliation{Institute of Laser Engineering, Osaka University, 2-6 Yamada-oka, Suita, Osaka, 565-0871, Japan}

\author{J.-R. Marqu\`es}
\affiliation{LULI, UMR 7605, CNRS, Ecole Polytechnique, CEA, Universit\'e Paris-Saclay, UPMC: Sorbonne Universit\'es, F-91128 Palaiseau cedex, France}

\author{A. Morace}
\affiliation{Institute of Laser Engineering, Osaka University, 2-6 Yamada-oka, Suita, Osaka, 565-0871, Japan}

\author{C. Moss\'e}
\affiliation{PIIM, Univ. Aix-Marseille - CNRS, France}

\author{O. Peyrusse}
\affiliation{PIIM, Univ. Aix-Marseille - CNRS, France}

\author{S. Rose}
\affiliation{Blackett Laboratory, Imperial College London, Prince Consort Road, London SW7 2BW, UK}

\author{M. Roth}
\affiliation{Institut f\"ur Kernphysik, Tech. Univ. Darmstadt, Germany}

\author{S. Sakata}
\affiliation{Institute of Laser Engineering, Osaka University, 2-6 Yamada-oka, Suita, Osaka, 565-0871, Japan}

\author{G. Schaumann}
\affiliation{Institut f\"ur Kernphysik, Tech. Univ. Darmstadt, Germany}

\author{F. Suzuki-Vidal}
\affiliation{Blackett Laboratory, Imperial College London, Prince Consort Road, London SW7 2BW, UK}

\author{V.T. Tikhonchuk}
\affiliation{Univ. Bordeaux, CNRS, CEA, CELIA (Centre Lasers Intenses et Applications), UMR 5107, F-33405 Talence, France}

\author{T. Toncian}
\affiliation{Institute for Radiation Physics, Helmholtz-Zentrum Dresden-Rossendorf e.V., 01328 Dresden, Germany}

\author{N. Woolsey}
\affiliation{Department of Physics, Heslington, University of York, YO10 5DD, UK}

\author{Z. Zhang}
\affiliation{Institute of Laser Engineering, Osaka University, 2-6 Yamada-oka, Suita, Osaka, 565-0871, Japan}

\date{\today}
\begin{abstract}
Powerful laser-plasma processes are explored to generate discharge currents of a few $100\,$kA in coil targets, yielding magnetostatic fields (B-fields) in excess of $0.5\,$kT. 
The quasi-static currents are provided from hot electron ejection from the laser-irradiated surface. According to our model, describing qualitatively the evolution of the discharge current, the major control parameter is the laser irradiance $I_{\mathrm{las}}\lambda_{\mathrm{las}}^2$.  
The space-time evolution of the B-fields is experimentally characterized by high-frequency bandwidth B-dot probes and by proton-deflectometry measurements. 
The magnetic pulses, of ns-scale, are long enough to magnetize secondary targets through resistive diffusion. We applied it in experiments of laser-generated relativistic electron transport into solid dielectric targets, yielding an unprecedented 5-fold enhancement of the energy-density flux at $60\mum$ depth, compared to unmagnetized transport conditions.
These studies pave the ground for magnetized high-energy density physics investigations, related to laser-generated secondary sources of radiation and/or high-energy particles and their transport, to high-gain fusion energy schemes and to laboratory astrophysics.
\end{abstract}

\maketitle

\section{Introduction}

Magnetic fields (B-fields) are ubiquitous in the Universe, where they rule many high-energy phenomena {\it e.g.}, magnetic arches (stellar flares, accretion disks, etc.), plasma jets and shock waves (supernova remnants, gamma-ray bursts, pulsar wind nebulae, etc.). In certain objects such as compact stars, these fields are so strong ($BÊ\simÊ10^4$-$10^5\,$T in white dwarfs, $\simÊ10^8$-$10^9\,$T in radio pulsars) that they determine the star's structure and composition, as well as its radiation properties\,\cite{Engel2008}. The reason is that, at the atomic level, the electron cyclotron energy is comparable to, or larger than the Coulomb binding energy. In other astrophysical settings, expanding plasma outflows can generate turbulent B-fields through collisionless shocks and/or magnetic reconnection mechanisms, leading to the production of high-energy particles and radiation\,\cite{Spitkovsky2008, Matsumoto2015}. 

Efforts in understanding these processes have up to now been restricted to a combination of modeling and observational analysis. Only recently has the unique potential of powerful lasers to reproduce similar physical conditions been fully realized and started being explored, thereby driving forward the relatively young field of laboratory astrophysics\,\cite{Remington2006, Fox2013, Huntington2015}.Ê

Besides astrophysical applications, there has been a growing interest over the past years in laser-driven, high-energy-density (HED) systems embedded in strong magnetic fields, with the aim of breakthrough advances in, {\it e.g.}, inertial confinement fusion (ICF)\,\cite{Chang2011, Strozzi2012, Perkins2013}, particle sources\,\cite{Macchi2013, Arefiev2016} or atomic physics\,\cite{Lai2001, Murdin2013}. The general goal is to produce B-fields strong enough that the associated field energy density is at least a significant fraction of the whole energy density, and/or the Larmor radius (respectively the cyclotron period) of some constituents becomes of the order of, or smaller than the relevant space (respectively time) scales of the problem.

In the framework of Inertial Confinement Fusion (ICF), controlled laser-driven implosions in magnetised conditions are a proposed strategy towards higher fusion gains\,\cite{Perkins2013}. It has already been demonstrated experimentally that imposed seed B-fields of $\sim10\,$T can be amplified by a $>500$ factor by field advection in a spherical implosion\,\cite{Chang2011}. These fields induce anisotropic thermal-electron diffusion and may reduce heat conduction to the dense core, therefore increasing implosion efficiency or even suppressing heat loss from the burning region once fusion reactions are initiated. The target may remain compressed over a longer time scale and with less stringent requirements on compression than in conventional inertial fusion experiments, as lower compression ratios tend to stabilize the imploding shell target. Implosions under magnetized conditions may also contribute to the reduction in the growth rate of hydrodynamic instabilities\,\cite{Sano2013}. The B-fields can also effectively confine the D-T ions and thermonuclear $\alpha$-particles enhancing collisionality and fusion yield. Thus, the study of highly magnetized plasmas is of critical importance.

The few laser experiments performed to date on externally magnetized samples have relied on capacitor-bank pulsed discharges in solenoids, {\it e.g.} \cite{Ciardi2013, Fiksel2014, Albertazzi2014, Revet2017}, with field strengths limited to $\sim 40\T$ ($20\T$ in regular operation)\,\cite{Pollock2006, Gotchev2009, Albertazzi2013}. In spite of assuring B-fields of interesting space- and time-scales, with cm-scale uniformity over of a few $100\,\mathrm{\mu s}$, the needed additional multi $10\,$kV electric pulse-power limit their easy deployment in any laser facility. Besides, the rather closed geometry of the pulser coil(s) chamber placed into the laser interaction chamber renders this technology cumbersome in experiments requiring wide access angles either for diagnostics, or for the magnetization and laser-driving of secondary samples.

This motivates the development of open-geometry, all-optical magnetic generators, portable to any high-energy laser facility. 
Our recent experiments pave the way to controlled laser-driven sources of strong quasi-static B-fields. More specifically, by driving capacitor-coil targets\,\cite{Daido1986} by ns, high-energy, tightly focused laser pulses, we have reproducibly produced magnetic pulses in excess of $0.5\,$kT over a few ns\,\cite{Santos2015, Law2016}. In this paper, Section \ref{Sec-B-fields} revisits the experimental characterization of laser-produced B-fields with a deeper insight into proton-deflectometry measurements and the difficulty of their interpretation. Further, we present new results of time-resolved optical shadowgraphy of the driven coils, revealing the development of a current-driven instability along the wire surface. The spatio-temporal scales of the excited surface mode provide a new way of estimating the discharge current in the coil.   
Such source of magnetostatic fields was readily applied to magnetize solid-density foils driven by an auxiliary intense laser, and we successfully demonstrated the radial confinement of laser-accelerated relativistic electron beams (REB) propagating in solid-density matter\,\cite{Bailly2017}. Section \ref{Sec-REB} summarizes the main results of our REB transport experiments and details on the mechanisms ruling the energy transport. A few perspective spinoffs of our platform are presented in Section \ref{Sec-Perspectives}.

\section{Laser-driven strong magnetostatic fields \label{Sec-B-fields}}

\begin{figure}
\center
\includegraphics[width=\columnwidth]{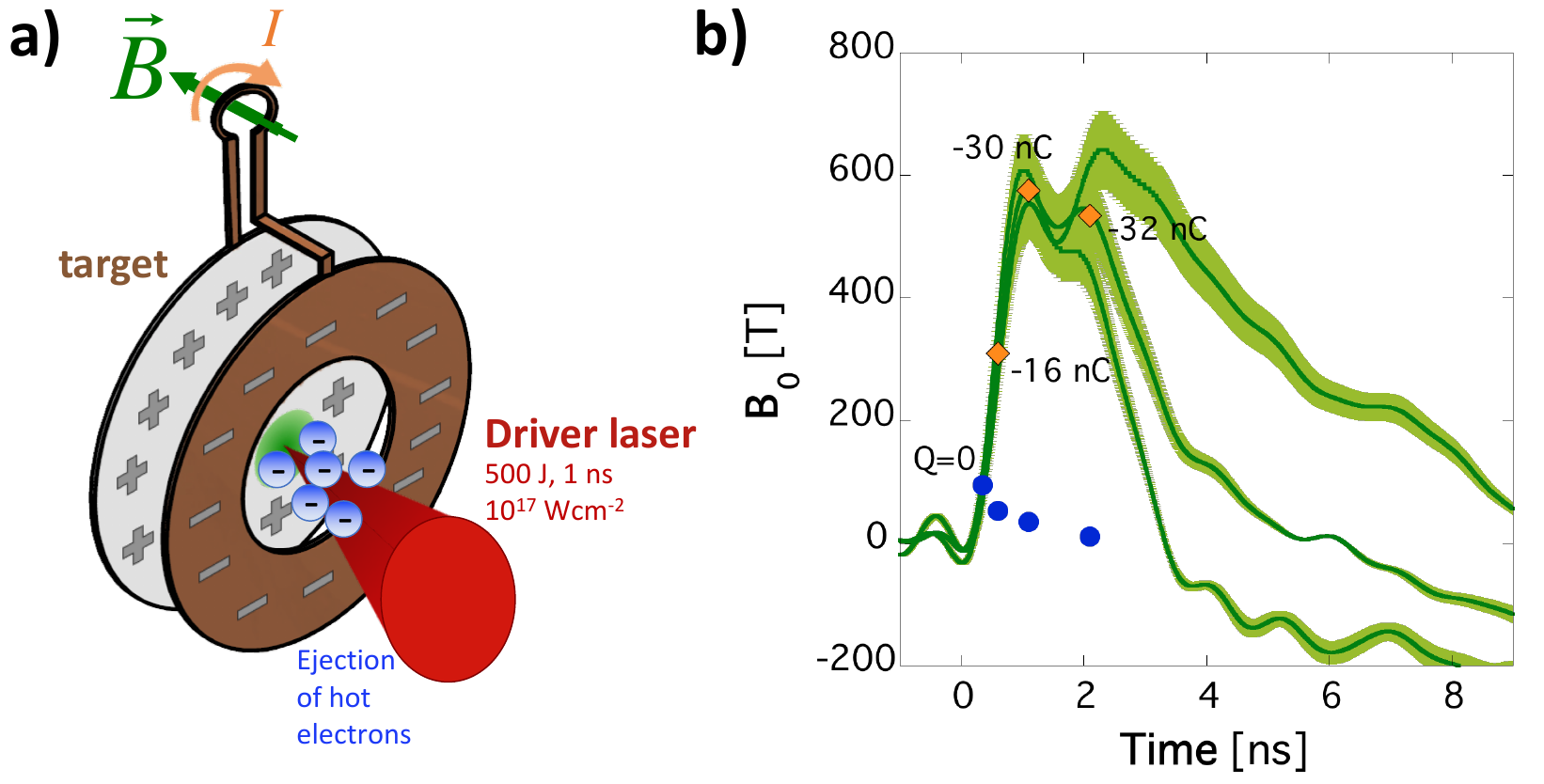}
\caption{\small {\bf a)} Illustration of the B-field production mechanism with laser-driven coil-targets: An intense ns laser ejects hot electrons, corresponding to a diode-like current source. The current loops in the wire, producing a dipolar-like B-field in the coil region. {\bf b)} B-field at the coil center as a function of time, estimated from B-dot probe measurements and 3D magnetostatic extrapolation (green curves) and from void bulb size in proton-deflectograms, compared to synthetic deflectograms assuming either a dipolar B-field of free strength [blue circles, see Fig.\,\ref{Fig-RCF}-c)], or adding an electrostatic charge of electrons $Q$ (labels) to the dipolar B-field of the strength obtained from the B-dot probes [orange diamonds, see Fig.\,\ref{Fig-RCF}-d)].  
\label{Fig-B-field}}
\end{figure} 

The production of strong magnetostatic fields by laser interaction with matter uses the target design depicted in Fig.\,\ref{Fig-B-field}-a): The so-called capacitor-coil targets are composed of two parallel disks at a distance $\sim 1\mm$ from each other, connected by a coil-shaped wire. 
The target charging results from the high-power ns laser passing through a hole on the front disk and interacting with the rear disk, creating a blow-off plasma and ejecting hot, supra-thermal electrons into the vacuum between the disks. 
Self-consistently, a quasi-static diode-like potential structure builds up across the disks, determining the maximum ejected electron current. 
Simultaneously, the coil-shaped wire reacts like an RL-circuit, giving rise to a looping discharge current $I$ and generating a dipole-like B-field $\vec B$ around the coil over a time-scale of a few ns. The B-field at the coilÕs center is approximately $B_0\approx \mu_0 I/2a$, where $\mu_0$ is the vacuum permeability and $a$ is the coil radius.
This scheme is a development of the design proposed back in the 80's\,\cite{Daido1986}, and recently explored by several groups to produce sub-kT B-fields\,\cite{Santos2015, Law2016, Fujioka2013, Goyon2017}.

In our experiments carried out at the LULI2000 laser facility (Ecole Polytechnique, France), we used $500\,$J energy, $1\,$ns square-long, $1.06\mum$-wavelength ($1\omega_0$) laser pulses, focused to $\sim10^{17}\Wcm$ intensities. 
The supra-thermal electron population was characterized by a $T_h\approx 40\pm 5\,$keV temperature, as measured by X-ray spectroscopy in the range of 10 to $1000\keV$. 
The temperature of the thermal electron component, $T_e\approx 1.2\pm 0.3\keV$, was characterized by Bragg diffraction spectroscopy in the range of $1.3$ to $1.7\keV$, coupled to atomic-physics calculations.
Under such conditions, peak B-field strengths $B_0 > 500\,$T were measured from targets with coil radius $a=500\mum$ [sample measurements in Fig.\,\ref{Fig-B-field}-b)], corresponding to peak currents of several $100\,$kA\,\cite{Santos2015}.

Owing to the very small capacitance of the disks, $C\sim 0.1\,$pF (geometrical details are given below), the target can be modeled as a laser-driven diode (the two disks) coupled with a $RL$ circuit (the coil-shaped wire)\,\cite{Tikhonchuk2017}. 
Electron pinching due to plasma self-generated B-fields limits the maximum current. This is mostly important at early times, for currents approaching the Alfv\'en limit, $17\,$kA. Then, the diode current is limited by the space charge.
As ions are electrostatically pulled by the expanding electron cloud, their inertia determines two different regimes for the diode: 
i) For times shorter than the ion-transit time between the plates ($\sim 100\ps$), the ejected electrons flow in vacuum, building up a $V\sim -10 T_h$ potential barrier which limits the current to $\sim 100\,$A. ii) Once the ions fill the volume between the disks at a density $\sim10^{18}\pcc$, the space charge is neutralized and the potential flattens, yielding a stationary electron flow. Adapted plasma and wire impedances allow intense currents, above $100\,$kA, just limited by the cathode potential jump, now reduced to $\vert V\vert \sim100\,$kV.
The current evolution $I(t)$ can be estimated from the wire equation $V=L\frac{\der I}{\der t} + R(t)I$ and considering the wire resistance $R(t)$ evolving according to the Joule heating. 
The current rises during the laser irradiation, while electrons keep on being accelerated and ejected, and then decays according to the circuit characteristic time, $\sim L/R$.

In summary, and according to our model detailed in Ref.\,\cite{Tikhonchuk2017}, three physical aspects are important to explain the intense discharge currents measured experimentally: the charge neutralization and the flattening of the potential distribution between the two plates, the ion inertia allowing for neutralized electron transport with currents far above the Alfv\'en limit, and the maximum temperature of the wire due to the latent heat of vaporization. The main control parameter is the hot electron temperature, which mainly ensues from the laser irradiance, $I_{\mathrm{las}}\lambda_{\mathrm{las}}^2$. Higher currents -- and stronger B-fields -- can be expected for high-intensity lasers at large wavelengths.

\subsection{B-field measurements}

\begin{figure*}
\center
\includegraphics[width=\textwidth]{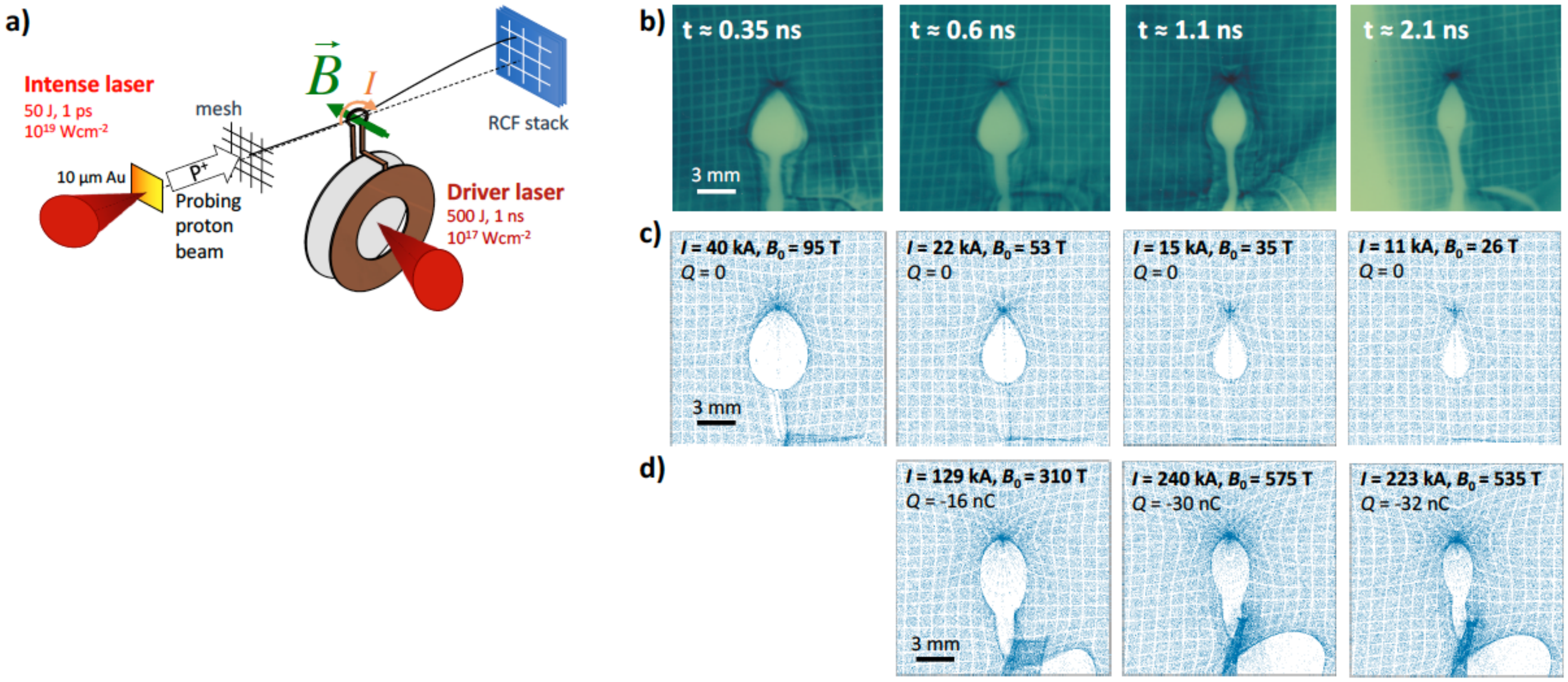}
\caption{\small {\bf a)} Sketch of the proton-deflectometry setup (not to scale). 
{\bf b)} Zoomed RCF data for $13\pm1\,$MeV protons from shots with varying delay between the lasers, $\Delta t = 0.25$, 0.5, 1 and $2\,$ns (from left to right). The labels give the probing times, accounting for the protons' time-of-flight between the foil and the coil. 
{\bf c)} Corresponding synthetic proton-deflectograms, obtained from numerical simulations of the proton trajectories through 3D B-field maps matching the experimental data (labelled by the discharge current $I$ and the corresponding strength of the B-field at the coil center $B_0$). 
{\bf d)} Idem setting the B-field strength to the average value inferred from the B-dot data at the corresponding times [orange diamonds in Fig.\,\ref{Fig-B-field}-b)] and adding a charge of magnetized plasma electrons $Q$, which allows matching the synthetic and the experimental widths of the void-bulb.
The white/black horizontal bars give the spatial scale at the RCF detection plane, corresponding to a magnification of 10 in respect to the plane of the coil axis.   
\label{Fig-RCF}}
\end{figure*} 

We performed the experimental characterization of the laser-driven B-fields at the LULI2000 facility\,\cite{Santos2015}.
Each of our capacitor-coil targets was made from one only laser-cut $50\mum$-thick metallic foil which was then fold to form two parallel disks, of $3500\mum$ diameter, connected by a coil-shaped wire of $50\mum$-side squared-section [see Fig.\,\ref{Fig-B-field}-a)]. The coil radius was $a=250\mum$ and the open angle between the coil's legs was $23.6\degree$. The coil center was at a height of $3\mm$ in respect to the center of the disks. The distance between the disks varied within $d=900\pm 200\mum$ due to the manual target folding process. In some cases, this also compromised the exact parallelism between the disks.
  
We used three independent diagnostics for the B-field measurements: 
i) High bandwidth probing of the time-derivative of the B-field (B-dot probe) at a few cm from the coil.
ii) Faraday rotation of the direction of polarisation of an optical probe laser through birefringent crystals placed at a few mm from the coil.
iii) Direct measurements of the B-field at the coil center were possible by proton-deflectometry. The probing proton beam was accelerated by an intense laser pulse interacting with a backlighter foil target. 

B-dot probes provide high temporal and spectral resolution and are practically insensitive to electric fields (E-fields). By contrast, they are extremely sensitive to B-fields, as low as a few $\unit{\mu T}$. The pickup coils should be placed at a few cm from the laser-driven coil, so that the B-field strength is below the probe safety threshold of $\sim 30\,$mT (at GHz frequency), and the B-field weakly varies over the cm-scale size of the probe. Within a single laser shot the probes follow the B-field evolution over hundreds of ns, with a resolution as good as $10\ps$. The B-field distribution in the region around the coil at each time is carefully extrapolated from the distant probe measurements by means of 3D magnetostatic simulations\,\cite{Radia}: the looping discharge current $  I$ is left as a free parameter in order to adjust the calculations to the measured B-field value\,\cite{Santos2015}. 

Faraday rotation is totally insensitive to E-fields. It uses birefringent crystals, the sensitivity of which depends on $\int_0^l Y \vec B\cdot d\vec z$, where $Y$ is the crystal's Verdet constant and $l$ its length in the propagation direction $z$ of the probing laser. Terbium gallium garnet (TGG) crystals with $Y=11.35\degree$T$^{-1}$mm$^{-1}$ can be quite small, $l<1\mm$, and can be positioned very close to the laser-driven coil. Unfortunately the optical performance may suffer from the laser-target interaction and the harsh plasma conditions. Also, it is reasonable to think that $Y$ may vary with the crystal temperature and be affected by the time-dependent character of the B-fields. Such dependencies are little known, and are possible sources of error in the evaluation of the B-field strength. Besides, in our experiments the birefringent crystals turned out to darken quite rapidly because of exposure to hard x-rays emitted from the laser interaction area.

Proton beams can directly probe the B-field distribution in the coil region. These beams were produced through target-normal sheath acceleration (TNSA) at the rear of thin foils irradiated by intense laser pulses\,\cite{Snavely2000}. Such sources have a broad energy spectrum, up to $\sim 20\MeV$ for our foil-targets and laser parameters. The deflections experienced by the protons through the B-field distribution were inferred from their imprints over a stack of multiple radiochromic films (RCF). 
The time resolution is guaranteed by the protons' different time-of-flight (TOF) between the back-lighter foil and the coil and by the protons' stopping power in matter, characterized by the very localized Bragg peak. This ensures that the most of a given proton energy is deposited at the end of its penetration range into the material. Protons of increasing energy bins are absorbed in the subsequent RCF layers of the detection stack. 
The images of the probed region of interest may have a spatial resolution as good as $\sim 10\mum$, with a time resolution better than $10\ps$. In one shot the covered time-range is limited to $\sim 100\ps$ (depending on the protons' energy dispersion and the different TOF). One needs multiple laser shots to cover the full time-range of the magnetic pulses -- of a few ns as revealed by the B-dot probe measurements -- varying the delay between the ns-laser driving the coil discharge and the ps-laser accelerating the probing proton beam. In practice, the experimental characterization of the B-field distribution is made difficult by the fact that the protons are also sensitive to E-fields and other plasma effects.

We used capacitor-coil targets of either Cu, Ni or Al\,\cite{Santos2015}. Here we focus on the results obtained with the more scrutinized Ni targets. 
Fig.\,\ref{Fig-B-field}-a) shows the evolution of the B-field at the coil center, $B_0$ (curves), as extrapolated from B-dot probe measurements. 
The peak B-field, $\approx 600\,$T (corresponding to a peak discharge current $I \approx 250\kA$), and the rise time, consistent with the $1\,$ns laser pulse duration, are reproducible. However, the decay time varies in a shot-to-shot basis in the range between 3 and $10\,$ns.

\subsubsection{Insight into proton-deflectometry measurements}

The B-field spatial distribution around the coil was measured by proton-deflectometry. 
The experimental setup is schematized in Fig.\,\ref{Fig-RCF}-a).
The probing protons were accelerated from $10\mum$-thick Au foils. The proton beam axis was set perpendicular to the coil axis. The distances from the Au foil to the coil and from the coil to the RCF stack used for proton detection were respectively set to $d=5\mm$ and $D=45\mm$, translating into a imaging magnification of $M=10$ from the plane of the coil axis into the proton imprint signals. In order to quantify proton deflections and characterize the B-field distribution up to several mm transverse distances from the coil, a metallic mesh of $42\mum$-pitch was positioned between the Au foil and the coil, at $2\mm$ from the former. The pitch of its projection to the plane of the coil center is $105\mum$. 

Figure \ref{Fig-RCF}-b) shows sample images of proton imprints from different shots with varying delay $\Delta t$ between the ps-laser accelerating the protons and the ns-laser driving the coil target. The shown RCF layers correspond to imprints of $13\pm 1\MeV$ protons, and their labels indicate the probing times relative to the beginning of the ns laser drive, taking account of the protons' TOF.
The B-field distributions are inferred from deformations of the mesh imprint, as well as from the size and shape of the central bulb, void of any proton as the more centered of the incident probing protons are expelled from the regions of stronger B-field. The bulbs' characteristic shape, looking like a pear, is due to the sens of the current looping in the coil with respect to the proton beam axis [see Fig.\,\ref{Fig-RCF}-a)]. The horizontal component of the Lorentz force due to the strong poloidal B-fields around the coil rod pushes protons inwards and outwards at respectively the coil's top and bottom parts.   

Figures \ref{Fig-RCF}-c) and d) show the synthetic counterparts of Fig.\,\ref{Fig-RCF}-b). These images result from simulations of the trajectories of randomly injected protons within the same energy bin of the experimental signals\,\cite{prorad}. The protons propagate through a 3D B-field map, obtained from magnetostatic calculations for our coil-target geometry.     
Results in Fig.\,\ref{Fig-RCF}-c) assume only B-field effects, and the coil discharge current $I$ is used as an adjustable parameter until the synthetic images match the experimental ones in terms of both mesh-imprint deformations and bulb shape and size. The agreement is remarkable.
As inferred from the decreasing bulb sizes and mesh deformations, the evaluated B-field strength would decrease with time [blue symbols in Fig.\,\ref{Fig-B-field}-b)], in contradiction with the B-dot probe signals (green curves) which increase until $t\sim 1\,$ns, consistently with the laser pulse duration. Indeed, the deduced values of $B_0$ are consistent with the B-dot probe results only up to $\approx 0.35\,$ns. The assumption that only the driven B-field acts on the proton-trajectories leads to an under-estimation of the B-field strength for later times.

The above inconsistency was understood as the effect of progressive magnetization plasma electrons, building up an electrostatic potential near the coil region. 
Indeed, by increasing the time delay between the ns and ps laser pulses, the deflectograms show increasing deflections of the relativistic electrons emitted from the backlighter target\,\cite{Santos2015}, along with the decreasing deflections observed for protons. The opposite evolution of the protons' and relativistic electrons' deflections was interpreted as the signature of continuous negative charge accumulation in the vicinity of the coil, due to the easy magnetization of the ns-laser created plasma electrons. 
This effect is modeled in Fig.\,\ref{Fig-RCF}-d) by adding to the B-field distribution the electrostatic field created by a negatively and uniformly charged sphere of radius $250\mum$ (corresponding to the radius of the coil), of variable total charge $Q$ and centered at $250\mum$ below the coil center (centered in the gap of the coil rod). This position was heuristically chosen upon the simulation of electron trajectories randomly injected from below the coil: trapped particles tended to converge into periodic orbits around this region. The B-field was set to  the strength inferred from the B-dot data at the corresponding times.
Even if the obtained bulb shapes do not exactly reproduce those on the experimental images (the bulb shape depends considerably on the position and the distribution of the charge $Q$), the results lead to the conclusion that a magnetized charge of $\sim -30\,$nC is already enough to yield comparable bulb sizes.

Moreover, in a subsequent experiment carried out at the Gekko XII - LFEX laser facility (ILE, Univ. Osaka, Japan), we added a metallic shield protecting the coil from the interaction region between the disks. Peak B-fields of $B_0=600\,$T at the center of Ni coil-targets were then consistently measured by B-dot probing and by proton-deflectometry, without the need to hypothesize electron accumulation in the proton sampling area\,\cite{Law2016}.

The proton deflection maps confirmed the dipole-like spatial distribution of the B-field around the coil and the spatial-integrated energy of the coil B-field at peak-time corresponds to 4.5\% of the driver laser energy.

\subsubsection{Coil expansion and MHD instabilities}

\begin{figure}
\center
\includegraphics[width=\columnwidth]{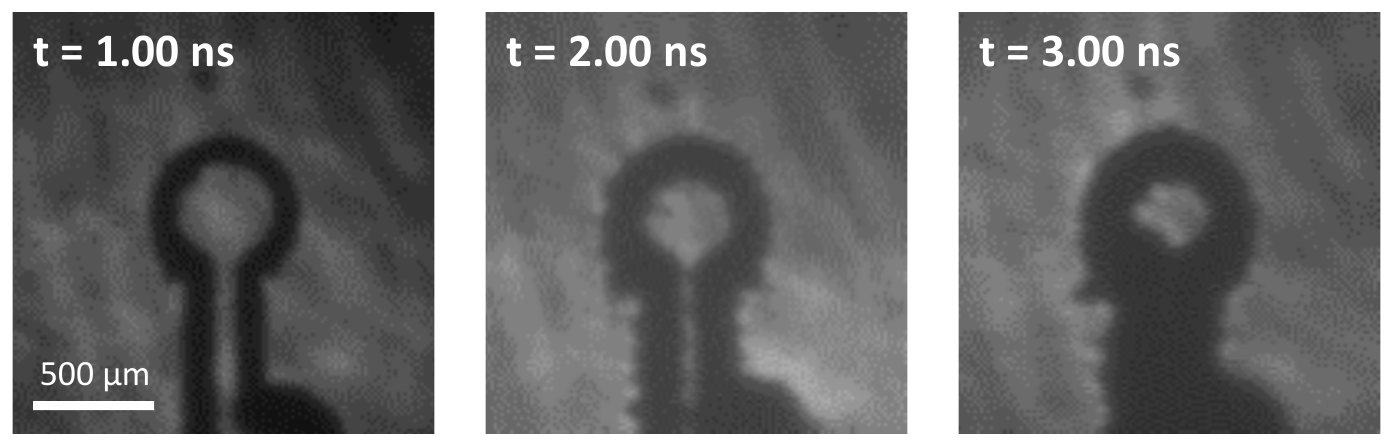}
\caption{\small Optical shadowgraphy (0.2-ns-gated images at $532\nm$) of the coil at three different times after laser driving of the capacitor-coil target. 
\label{Fig-shadow}}
\end{figure}

The strong electron current flowing in the wire leads to the Ohmic heating and melting of its surface, and may trigger a magnetohydrodynamic (MHD)-type instability.
Figure\,\ref{Fig-shadow} shows $0.2\,$ns-gated shadowgraphic images of Ni-coils at three different times after the start of the laser drive. One can clearly see the wire expansion at $v_{\mathrm{wire}}\sim 10\,\mathrm{\mum/ns}$, and also a periodic transverse modulation of its surface with a characteristic wavelength $\lambda_{\mathrm{wire}}\approx 110\pm 10\mum$. The modulation seems to grow from $t>1\,$ns -- once the discharge current reaches its maximum [see Fig.\,\ref{Fig-B-field}-b)] -- at a rate of $\sim 10^9\,$s$^{-1}$.

Such an instability is probably due to the competition between the thermal and the magnetic pressures at the wire surface, similar to the interchange or sausage instability in z-pinches\,\cite{Greene1968}. It is excited if the metal is in a plasma state and the magnetic pressure at the wire surface is larger than the thermal pressure. The growth rate of this instability is $\gamma \simeq kv_{\mathrm{A}}$, with the modulation wave vector $\vec k$ parallel to the wire axis and the Alfv\'en speed $v_{\mathrm{A}}= \sqrt{ B_\varphi^2/\mu_0 \rho}$. Here $B_\varphi$ is the value of the poloidal B-field close to the Ni wire surface, of mass-density $\rho \approx 9\gcc$. According to the observations in Fig.\,\ref{Fig-shadow}, the growth-rate is at least $\gamma \sim 1\,$ns$^{-1}$, and the wave number  $k\sim2\pi/\lambda_{\mathrm{wire}}\sim 570\,$cm$^{-1}$. Comparing these data with the formula for the growth rate, one can estimate a value for the poloidal B-field required for the instability development, which is $B_\varphi\sim1800\T$. Remarkably, this is consistent with a value of the poloidal B-field at the wire surface yielded by the peak current $I\approx 250\kA$ inferred from the analysis of the B-dot probe data.   
  
Given the open geometry and the relatively small expansion velocity of the coil's rod, the B-field generator under consideration can be readily exploited to magnetize secondary laser-driven samples. This all-optical experimental design offers the advantage of not relying on additional power discharge sources.

\section{Guiding of relativistic electron beams \label{Sec-REB}}

\begin{figure*}
\center
\includegraphics[width=\textwidth]{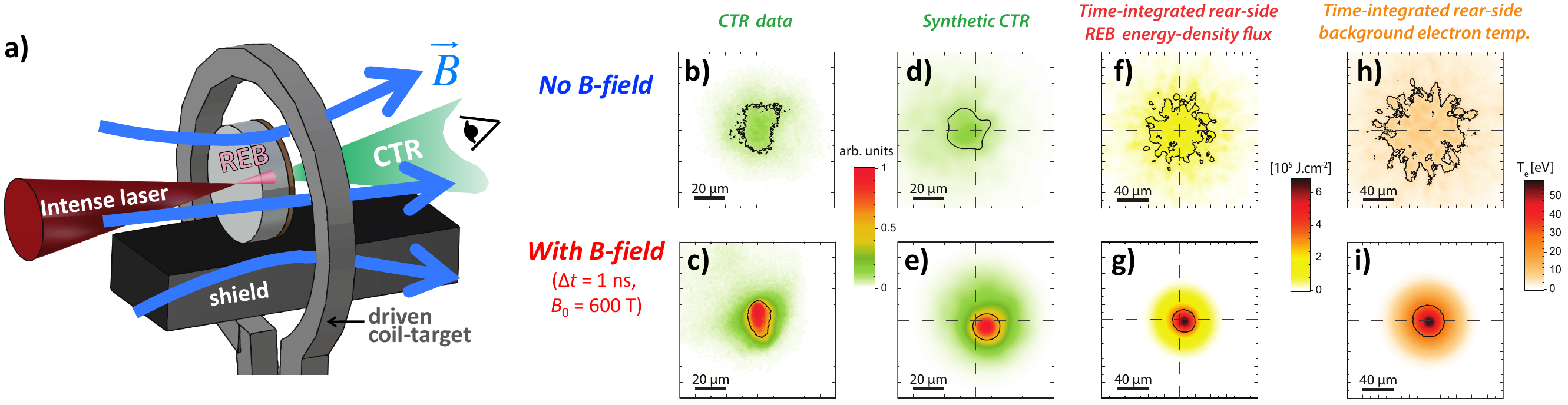}
\caption{\small {\bf a)} Sketch of the experimental setup for the REB-transport with imposed B-field. {\bf b)}, {\bf c)} CTR data b) without and c) with B-field.  {\bf d)}, {\bf e)} Synthetic CTR images d) without and e) with B-field. {\bf f)}, {\bf g)} Unfolded time-integrated REB energy-density flux at the targets' rear surface f) without and g) with B-field. {\bf h)}, {\bf i)} Final background electron temperature at the targets' rear surface h) without and i) with B-field. 
The black contour lines stand for the half-height of the signals and the dashed crosses give the position of REB-injection at the targets' front surface.
\label{Fig-REB}}
\end{figure*} 

We applied the above laser-driven B-fields to the guiding of relativistic electron beam (REB) through solid targets. 
For that, the transport targets -- of $50\mum$ CH with $10\mum$ Cu coating on the rear side -- were placed at the coil vicinity, as schematized in Fig.\,\ref{Fig-REB}-a). 
The REB were accelerated by the intense ps laser [$1\ps$ full-width at half-maximum (FWHM), $50\J$, $1.5\times 10^{19}\Wcm$] at different delays $\Delta t$ with respect to the ns laser driving the coils.
Besides scanning the laser pulses delay, we also tried different positions of the transport target in respect to the coil in order to test different target magnetization configurations\,\cite{Bailly2017}. Yet here we focus on the results obtained with the transport targets positioned at the coil plane, which assured an approximately radially symmetric distribution of the B-field embedded into the target, with respect to the axis of REB injection. 
It takes about $1\,$ns for the CH layers to be entirely magnetized, as estimated from the B-field diffusion time $\tau_{\mathrm{diff}} = \mu_0 L^2 / \eta \approx 1 \unit{ns}$ over the CH layer length $L=50\mum$, assuming a constant resistivity of $\eta=10^{-6}\,\mathrm{\Omega m}$. This simple estimate is supported by simulations of the B-field resistive diffusion (see Supplementary Information in Ref.\,\cite{Bailly2017}).
Therefore, for laser delays $\geq 1\,$ns, the REB experiences an essentially longitudinal B-field distribution close to that induced in vacuum, with a peak strength of $600\T$ weakly varying over the target thickness.

The transverse profile of the REB at the target rear side was investigated by imaging the $2\omega_0$ coherent transition radiation (CTR) emitted when the electron beam crosses the target-vacuum boundary. 
Figure Fig.\,\ref{Fig-REB}-b) and c) are sample CTR images, respectively without and with an externally applied B-field. A characteristic symmetric pattern\,\cite{Santos2002} is seen without the B-field. 
When imposing the B-field and shooting the intense laser at $\Delta t=1\,$ns, the 8 times higher yield and smaller size of the CTR signal reveal a radially-pinched electron beam propagation across the transport target.

In order to unfold the mechanisms of REB transport we reproduced the experiment in 3D PIC-hybrid simulations, with and without imposed B-field. 
The initial REB total kinetic energy was set to $30\%$ of the on-target ps-laser energy, and injected at the front surface over a region of $r_0\approx 25\mum$-radius half-width at half-maximum (HWHM). The injected electron kinetic energy spectra were characterized by power laws for the low energy part $\propto \varepsilon_{\mathrm{k}}^{-1.6}$ and exponential laws for the high energy part $\propto \exp \left( - \varepsilon_{\mathrm{k}} / T_{\mathrm{h}} \right)$ with $T_{\mathrm{h}}=1.3\,$MeV, as predicted by the ponderomotive potential. The injected angular distribution was characterized by a $30 \degree$ mean divergence angle and a $55\degree$ dispersion angle as defined in\,\cite{Debayle2010}. The total simulation time was set to $3.6\unit{ps}$ (with $t=1.25\unit{ps}$ corresponding to the peak REB flux at the front surface).

The simulation results were post-processed to obtain synthetic CTR signals.
The experimental CTR patternÕs size, shape and relative yield variations from unmagnetized to magnetized transport were fairly reproduced over the range of laser-target parameters considered, as reported in our previous reference paper\,\cite{Bailly2017}, yet, as mentioned above, here we restrict the discussion to the results obtained when the target is placed in the coil plane. 
Synthetic CTR images without B-field [Fig.\,\ref{Fig-REB}-d)] and with a $B_0=600\T$ B-field [Fig.\,\ref{Fig-REB}-e)] fairly agree with the experimental data: the simulations reproduce the experimental ratio of CTR yield (with B-field / without B-field) with $15 \pm 2\%$ relative errors, and the experimental CTR spot radius (azimuthally averaged), with or without B-field, with $15 \pm 5\%$ relative errors. 
Additional simulations show that the REB's radial confinement actually sets in above the threshold $B_0 \sim 400\T$, which corresponds to an electron Larmor radius for the hot electrons' average energy ($\sim 1\,$MeV) smaller than the REB source radius, $r_0\approx 25\mum$. By confronting the experimental and numerical results with and without the applied B-field, we found that our data are consistent with B-field strengths of $B_0\sim 500$-$600\,\rm T$\,\cite{Bailly2017}. Albeit indirect, this is consistent with our characterization of the generated B-field presented in Section \ref{Sec-B-fields}.    

\subsection{Evolution of the REB profile and transported energy}

\begin{figure*}
\center
\includegraphics[width=\textwidth]{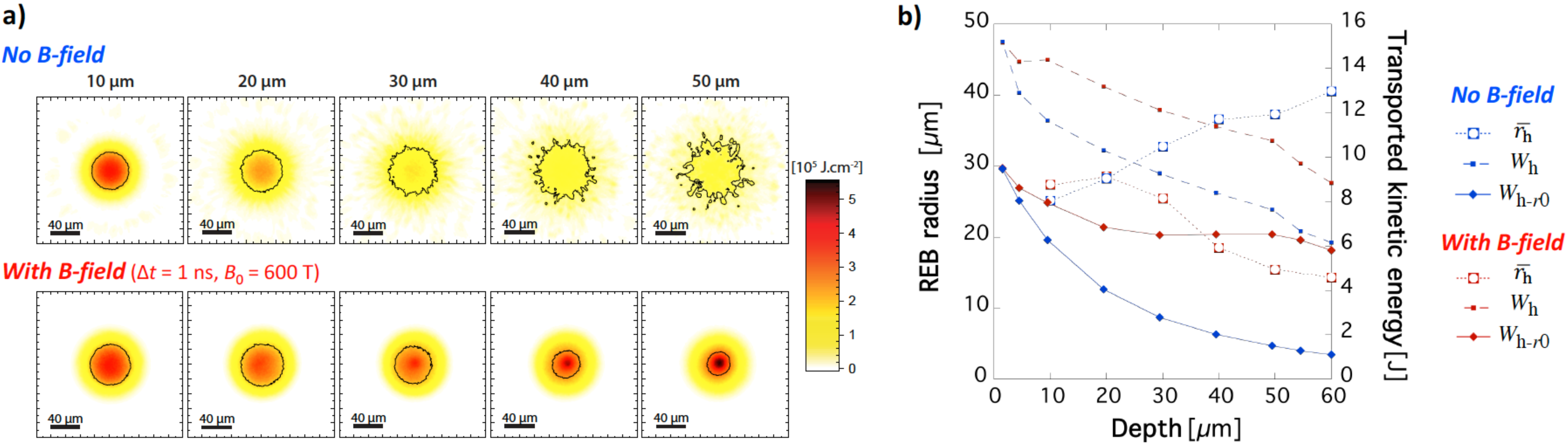}
\caption{\small {\bf a)} Sliced patterns of the time-integrated REB energy-density flux at different target depths, without (top images) and with (bottom images) imposed B-field. The black contour lines stand for the half-height of the signals. {\bf b)} Evolution along the propagation depth of the REB azimuthally-averaged radius, $\bar r_h$ (empty symbols, left-hand side ordinates), and transported kinetic energy (right-hand side ordinates) -- where $W_\mathrm{h}$ and $W_{\mathrm{h-}r_0}$ are respectively the total transported kinetic energy (full squares) and its fraction within the initial REB radius $r_0$ centered at the target axis (full diamonds) -- with (red) and without (blue) imposed B-field.
\label{Fig-REBtransport}}
\end{figure*} 

Figure \ref{Fig-REBtransport}-a) shows the transverse patterns of the time-integrated REB energy-density flux, without (top) and with (bottom) external B-field ($B_0=600\T$), for different depths into the transport target CH-layer.  
As expected\,\cite{Gremillet2002}, the electron beam undergoes strong filamentation when propagating in unmagnetized plastic, and the transported energy significantly spreads radially due to the intrinsic divergence of the REB source\,\cite{Debayle2010, Adam2006} and the collisional diffusion with the background ions.
In magnetized targets, by contrast, the REB-filaments are smoothed as the electrons are trapped and flow gyrating around the B-field lines. The B-field is actually strong enough to radially pinch the relativistic electrons. 

Figure \ref{Fig-REBtransport}-b) plots as a function of the target depth the azimuthally averaged radius (HWHM) of the REB energy-density flux patterns $\bar r_h$ (open symbols connected by dotted lines, left-hand side ordinates), and the time-integrated transported kinetic energy (right-hand side ordinates), both the total kinetic energy ($W_{\mathrm{k}}$, full squares connected by dashed lines) and the kinetic energy encircled over the surface corresponding to the initial REB-source, $\pi r_0^2$, kept centered with the injection axis ($W_{\mathrm{k-}r_0}$, full diamonds connected by solid lines). 
Ensuing from the diverging versus confined electron transport, about 45\% more energy is transported to the target rear in the magnetized case as a result of the magnetically confined low-energy electrons.
Much more importantly, from the efficient confinement results that the $r_0$-encircled energy $W_{\mathrm{k-}r_0}$ in the magnetized case contains $\approx 66\%$ of the REB total energy transported to the target rear, against $\approx 18\%$ for the unmagnetized case. 

In conclusion, the externally imposed B-field increases the time-integrated REB energy-density flux after the $60\mum$ target length by a $\approx 5.3 \times$ factor, as seen from the comparaison of Fig.\,\ref{Fig-REB}-f) and g). As a consequence the final background electron temperature rises to $\approx 60\eV$[Fig.\,\ref{Fig-REB}-i)], corresponding to $\sim 1 \unit{eV}$ per joule of laser energy at a $60\mum$ depth.
The reached temperature is a factor 5.9 higher than in the unmagnetized case [Fig.\,\ref{Fig-REB}-h)].  
Such unprecedented improvements of energy-density flux through dense matter and of the induced isochoric heating pave the way for advancing investigations in laser-driven sources of particles and radiation, in laser-driven thermonuclear fusion, and about matter states relevant for planetary or stellar science.

\section{Further perspectives in high energy-density physics \label{Sec-Perspectives}}

\subsection{Enhanced proton acceleration from solid targets}

\begin{figure}
\centering 
\includegraphics[width=0.6\columnwidth]{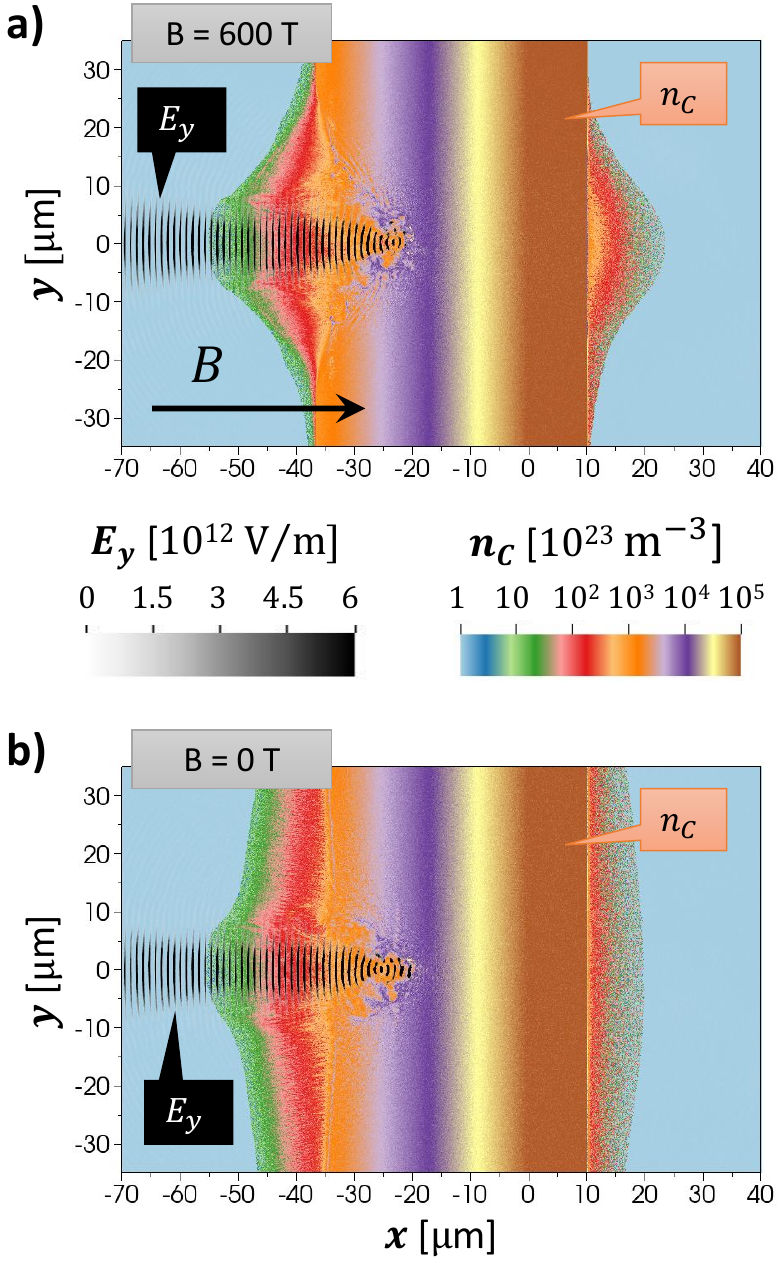}
\caption{Snapshots of carbon density $n_C$ (color scale) from PIC simulations of a CH target irradiated by an intense laser pulse, {\bf a)} with and {\bf b)} without an applied $600\,$T B-field. The y-component of the laser E-field is superposed (gray scale). The snapshots are taken $70\,$fs after the peak laser intensity.}
\label{Fig-density}
\end{figure}

The efficiency of production and transport of the laser-driven REB closely affects the optimization of laser-driven secondary particle sources, such as ion, positron, and even neutron sources. The already demonstrated ability to impact electron transport with an externally generated magnetic field suggests that the B-field may also be leveraged as an extra ``control knob'' for secondary particle sources\,\cite{Arefiev2016}. Here we focus on the possible impact of an externally applied $600\,$T field in laser-driven ion acceleration.

Again we consider a dielectric CH target, allowing for rapid B-field soaking, and laser parameters similar to those of the REB transport experiment. The following preliminary analysis was performed using fully kinetic 2D PIC simulations with the code EPOCH\,\cite{Arber2015}. Figure \ref{Fig-density} presents C-density maps from two runs, respectively a) with and b) without an imposed uniform and static B-field of $600\T$, directed normal to the laser-irradiated target surface.  
The laser pulse propagates along the $x$-axis. The laser focal plane in the absence of a target is located at $x = 0$. The FWHM laser spot size and duration are $5\mum$ and $900\fs$ respectively, yielding a $10^{19}\Wcm$ peak intensity in the focal plane. The pulse is linearly polarized, with the E-field in the $(x,y)$-plane of the simulations. In order to account for a laser prepulse, we introduce an extended preplasma at the laser-irradiated side of the target. The target material is fully ionized CH that we approximate by carbon ions and protons of equal density, $n_C = n_p = 10 n_{\mathrm{crit}}$ for $0 \leq x \leq 10\mum$ and $n_C = n_p = n_{\max} \exp(x / l)$ for $x < 0\mum$, where $l=8\mum$ and $n_{\mathrm{crit}}$ is the critical density for the $1.06\mum$ laser wavelength. The initial electron density is $n_e = 6n_C + n_p$. The simulation box width (along $y$) is limited to $80\mum$ due to computational time constraints.

The relativistically hot electrons generated in the preplasma up to the critical surface set up a sheath electric field wherever there is a density gradient. This is the field that then accelerates protons and carbon ions. 
The two $n_C$ snapshots in Fig.\,\ref{Fig-density} were taken $70\fs$ after the laser peak intensity. The impact of the applied B-field on the plasma expansion dynamics is already evident at both the front and rear sides of the target, causing the density change to be localized around $|y|< 30\mum$ [Fig.\,\ref{Fig-density}-a)], as a direct result of the confinement of the laser-accelerated relativistic electrons.  By contrast, in the absence of the B-field [Fig.\,\ref{Fig-density}-b)] the relativistic electrons quickly spread laterally and set up a much wider expansion at the rear surface.

\begin{figure}
\centering 
\includegraphics[width=0.7\columnwidth]{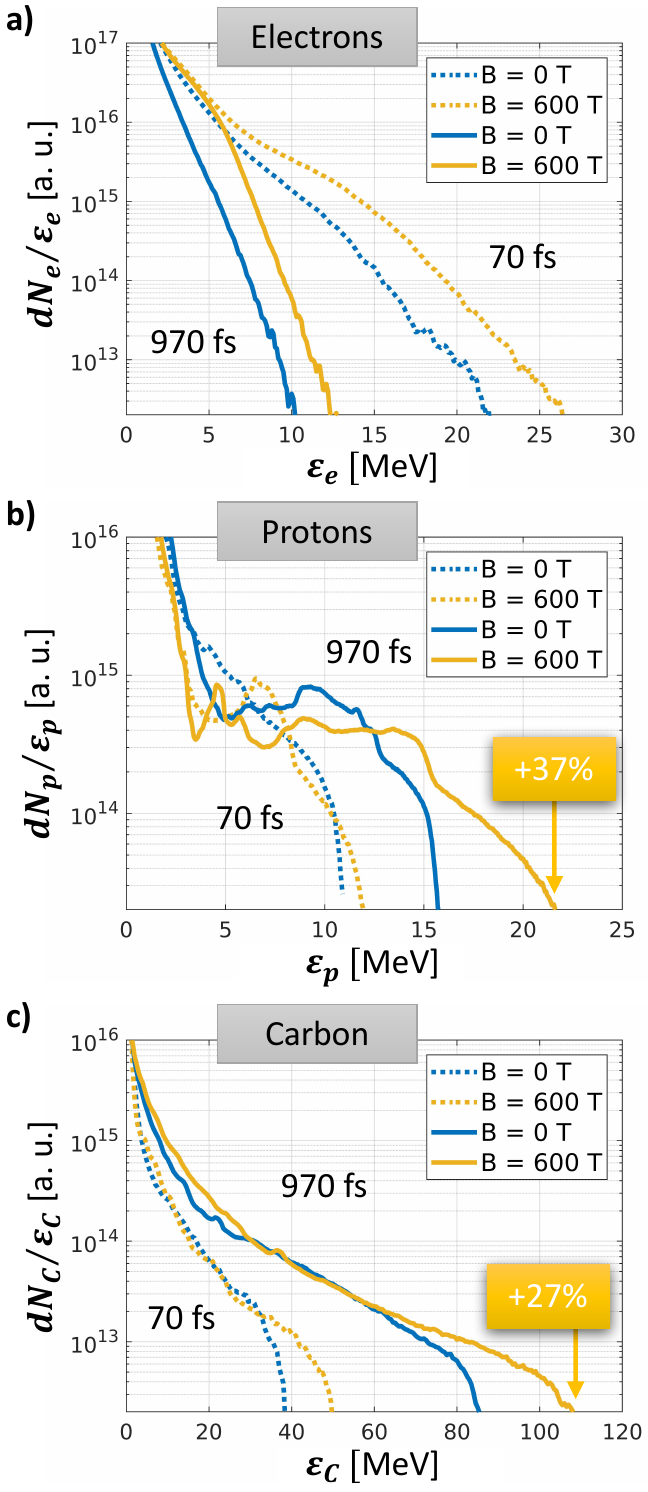}
\caption{Snapshots of energy spectra for {\bf a)} electrons, {\bf b)} protons, and {\bf c)} carbon ions in a laser-irradiate target with (yellow curves) and without (blue curves) an applied B-field. The snapshots are taken at $70\fs$ (dotted curves) and at $970\fs$ (solid curves) after the peak laser intensity. The shown spectra correspond to all the particles in the simulation at the given time.}
\label{Fig-spectra}
\end{figure}

Electron spectra are shown in Fig.~\ref{Fig-spectra}-a) at $70\fs$ (dotted curves) and at $970\fs$ (solid curves). In the case of the applied B-field (yellow curves) the electron spectrum is somewhat enhanced. It is worth noting that previous simulations with a much shorter laser pulse and a much shorter preplasma showed no significant impact on electron heating by an applied $1.5\kT$ B-field~\cite{Arefiev2016}. It remains to be determined using detailed electron tracking whether the changes in the electron spectra are caused by enhanced electron acceleration in the preplasma. There are multiple mechanisms that can potentially be impacted~\cite{Arefiev_JPP2015, Arefiev_PoP2016}, as the applied magnetic field changes transverse electron motion. We used open boundary conditions in the simulations in order to prevent artificial electron confinement in the lateral direction. Electrons leaving the simulation box while diffusing laterally through the target can be another contributing factor to the observed difference in the electron spectra. 

The enhancement of the electron spectrum and the lateral localization of the hot electrons combine to make the sheath field that accelerates ions stronger. A signature of this is a 27\% increase of the cutoff energy of the carbon ions already at $70\fs$ [Fig.~\ref{Fig-spectra}-c)]. However, the proton spectra at $70\fs$ show no such enhancement [dotted curves in Fig.~\ref{Fig-spectra}-b)]. This may indicate that the difference in strength of the sheath field builds up gradually, so that the leading edge protons quickly accelerate due to their greater mobility (compared to C) in a field that is still not impacted by the applied B-field.

The laser no longer heats the electrons $900\fs$ later. The electrons transfer the energy that they have accumulated interacting with the laser to the ions through the expanding sheath electric field. This is the cause for the decrease in the electron energy spectrum and the increase in both proton and carbon spectra. The relative enhancement of the cutoff energy in the carbon spectra remains unchanged at 27\%, while the absolute value more than doubles over $900\fs$. The proton spectra also experience a significant cutoff energy enhancement over the $900\fs$, of about 37\%, due to the applied B-field. 

We infer that a $600\T$ B-field can induce an appreciable enhancement of energetic ion spectra in laser-driven experiments with relativistic intensity ps-long laser pulses. 
We expect that this effect will be even more pronounced in a more realistic 3D numerical setup. The extra dimension would only increase the outflow of energetic electrons from the focal spot and suppress ion acceleration in the absence of an applied magnetic field.
Note that the present simulations neglect collisional processes. It remains to be determined whether their influence on hot-electron dynamics, namely through the driven resistive fields, can affect the ion acceleration in presence of an externally applied strong B-field.

\subsection{Magnetized atomic physics}

In certain settings, like in the crusts of white dwarfs or neutron stars, the ambient B-field is so important that it strongly influences the atomic physics processes and consequently the starÕs properties. 
The problem of atomic structure in the presence of a strong B-field is also of major fundamental interest\,\cite{Lai2001}. Under such conditions electrons are more efficiently bound to the nucleus along the direction of the B-field and the atom loses the spherical symmetry. Calculations are complicated because the diamagnetic term of the Hamiltonian has to be taken into account, which makes the Hamiltonian non-separable (no analytical solution even for hydrogen). This long-standing problem has been addressed through various theoretical approaches, but none of these have been benchmarked by laboratory measurements. 

Magnetic effects become strong in a given atomic shell when the electron cyclotron energy exceeds the shell's binding energy. Assuming a hydrogen-like atom, this means $B > 2.35\times10^5 Z^2/n^2\T$ (where $Z$ is the atomic number and $n$ the shell's quantum level). To create and diagnose such so-called Landau quantization states in the laboratory, we aim at performing absorption spectroscopy measurements of H and/or He atoms immersed in a sub-kT B-field. The atoms could be delivered as a gas jet (non-ionized, at least partially). For such light atoms, the spectral region of interest is 10 to $25\eV$. We may need a gas areal-density of $10^{15}\,\mathrm{cm^{-2}}$ to detect the signatures of the B-field effect on an absorption spectrum. 
  
\begin{figure*}
\center
\includegraphics[width=\textwidth]{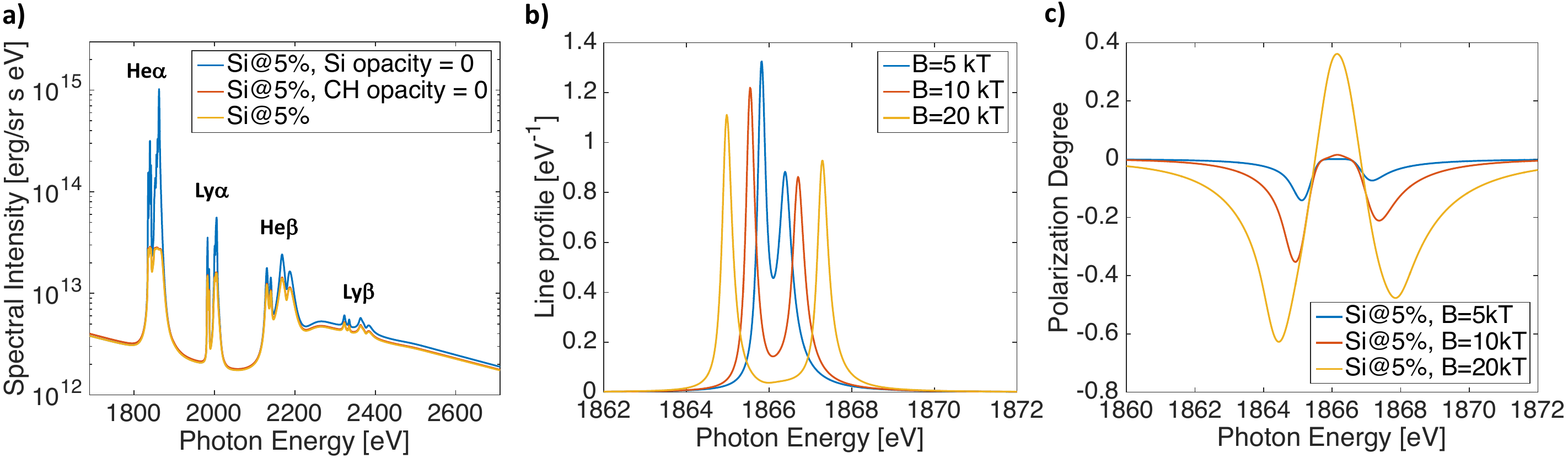}
\caption{\small {\bf a)} Calculated spectral intensity of Si (5\% atomic dopant) in a compressed CH plasma of $T_e\sim500\eV$ and density $n_e\sim 10^{23}\pcc$, including radiation-transport and opacity effects. {\bf b)} Stark-Zeeman effects on the $\sigma$-component of the Si He-$\alpha$ line emission and {\bf c)} polarization degree of the Si He-$\alpha$ line for different B-field strengths. The codes ABAKO, PPP-B and MASC were used.
\label{Fig-atom}}
\end{figure*}  

A different topic concerns the characterization of plasmas for which B-field effects can still be considered within a perturbative approach ({\it e.g.}, Zeeman effect), yet of the same order as collective plasma fields ({\it e.g.}, Stark effect). For highly charged ions in plasmas immersed in a B-field sufficiently strong that the coupling of the B-field to the atomic magnetic moment dominates the spin-orbit interaction, calculations predict line emission broadening and polarisation relative to the direction of the applied B-field: the emission $\sigma$-component is highly sensitive to the B-field. The shape of the C-VI emission lines (in the 350-$400\eV$ range) of a plastic target should provide interesting signatures of the kT-level B-field.
For B-fields in excess of $3\kT$, higher-Z elements (F, Al, or Si) can give access to B-field-modified spectral data in the keV-range. 

To study experimentally these effects we aim at producing dense plasmas from laser-driven implosions in the presence of frozen-in seed B-fields. The B-fields, generated by our all-optical platform, will be amplified through plasma compression to the multiple kT-range levels required for spectroscopic purposes. 
We consider here a compressed CH plasma of electron temperature $T_e\sim500\eV$ and density $n_e\sim 10^{23}$, with 5\% of doping Si atoms inserted as tracers of the magnetization level. We are using the atomic kinetics and radiation transport code ABAKO\,\cite{Florido2009} and the line shape code PPP-B\,\cite{Calisti1990, Ferri2011} coupled to the atomic physics code MASC to predict the Stark-broadened K-shell emission spectra of Si. The calculations in Fig.\,\ref{Fig-atom}-a) show positive evidence of tracer emission detectability while indicating a significant self-absorption effect on He-$\alpha$ and Ly-$\alpha$ lines. Oppositely, self-absorption is much smaller in $\beta$-type lines. 
The simulations of the Stark-Zeeman effect on the Si He-$\alpha$ line in Fig.\,\ref{Fig-atom}-b), convoluted with an instrumental broadening function of $0.25\eV$ FWHM ($E/\Delta E \approx 7500$) -- well in line with nowadays state-of-art x-ray spectrometers\,\cite{Beiersdorfer2016} -- show that interesting signatures are detectable for B-field strengths between 5 and $10\kT$, which should be reachable by advection in imploding plasmas. 
By recording simultaneously the $\sigma$- and $\pi$-components at a closer view and equivalent solid angles (two crystals oriented perpendicularly to each other) it is possible to characterize the polarization degree of the different Stark-Zeeman emission lines, $P=(I_{\pi}-I_{\sigma})/(I_{\pi}+I_{\sigma})$, with $I_{\pi}$ and $I_{\sigma}$ the intensity of the $\pi$- and $\sigma$-components of a given line. Such polarisation degree is calculated in Fig.\,\ref{Fig-atom}-c) for the Si He-$\alpha$ emission and different B-field strengths, using now an $1\eV$ FWHM instrumental function. 
Although degraded, this spectral resolution should be good enough to detect B-field-induced polarization effects for $B>5\,$kT. Several lines can be measured simultaneously to uniquely unfold B-field, electron density and temperature values. 

Such plasma characterization and increased understanding of the spectral properties of magnetized atoms in controlled laboratory samples would be an important contribution to fundamental science and of pivotal interest in magnetized ICF studies or in experiments of laboratory astrophysics.

\section*{Acknowledgments}

The research was carried out within the framework of the "Investments for the future" program IdEx Bordeaux LAPHIA (ANR-10-IDEX-03-02) and of the EUROfusion Consortium and has received funding from the Euratom research and training programs 2014-2018 under grant agreement No 633053 and 2017-2018 under grant agreement No CfP-AWP17-IFE-CEA-02. The views and opinions expressed herein do not necessarily reflect those of the European Commission. 
Part of the used diagnostic equipment was funded by the French National Agency for Research (ANR) and the competitiveness cluster Alpha - Route des Lasers, project number TERRE ANR-2011-BS04-014. 
The authors also acknowledge support from the COST Action MP1208 "Developing the physics and the scientific community for Inertial Fusion". 
We gratefully acknowledge the support of the LULI pico 2000 at Ecole Polytechnique and the Gekko XII at Univ. Osaka staff during the experimental runs. 
Simulation work on proton-deflectometry was partially supported by the Russian Foundation for Basic Research (contract 16-52-50019).
We also acknowledge the use of the PRORAD code, written by M. Black and maintained by S. Wilks.
Simulation work on relativistic electron transport has been partially supported by the Spanish Ministry of Economy and Competitiveness (grant No. ENE2014-54960-R) and used HPC resources and technical assistance from BCS and CeSViMa centers of the Spanish Supercomputing Network. 
Simulation work on ion acceleration was supported by the US Department of Energy under contract No. DE-SC0018312. Simulations were performed with EPOCH (developed under UK EPSRC grants EP/G054940, EP/G055165 and EP/G056803) using HPC resources provided by the TACC at The University of Texas at Austin.
Simulation work on atomic kinetics and radiation transport has been partially supported by Research Grant No. ENE2015-67561-R (MINECO/FEDER, UE) from the Spanish Ministry of Economy and Competitiveness.
The Japanese collaborators were supported by the Japanese Ministry of Education, Science, Sports, and Culture through Grants-in-Aid for Young Scientists (Grants No. 24684044), Grants-in-Aid for Fellows by JSPS (Grant No. 14J06592), and the program for promoting the enhancement of research universities.

\end{document}